\documentclass[english,draft]{aipproc}
\usepackage[T1]{fontenc}
\usepackage[latin1]{inputenc}
\usepackage{graphicx}
\usepackage{amssymb}

\makeatletter

\layoutstyle{6x9}

\usepackage{babel}
\makeatother
\begin{document}

\newcommand{\missET}{{E_{T}\!\!\!\!\!\!\!/\:\:\,}}

\newcommand\gsim{\mathrel{\rlap{\raise.4ex\hbox{$>$}} {\lower.6ex\hbox{$\sim$}}}}   \newcommand\lsim{\mathrel{\rlap{\raise.4ex\hbox{$<$}} {\lower.6ex\hbox{$\sim$}}}}   

\def\gtrsim{\gsim} 

\def\lesssim{\lsim} 

\def\alphas{\alpha_s }

\def\D0{D\O~} 



\author{Stefan Berge}{ address={Department of Physics, Southern Methodist University, Dallas, Texas 75275-0175, U.S.A.} }  \author{Pavel M. Nadolsky}{ address={High Energy Physics Division, Argonne National Laboratory,\\ Argonne, IL 60439-4815, U.S.A.} }  \author{Fredrick I. Olness}{ address={Department of Physics, Southern Methodist University, Dallas, Texas 75275-0175, U.S.A.} }  \author{C.-P.~Yuan}{ address={Michigan State University, Department of Physics and Astronomy, East Lansing, MI 48824-1116, U.S.A.} } 

\begin{flushright} ANL-HEP-PR-05-67  \\ hep-ph/0508215 \end{flushright}

\title{$q_{T}$ Uncertainties for $W$ and $Z$ Production\thanks{Contribution to the proceedings of the XIII International Workshop on Deep Inelastic Scattering (DIS 2005), April 27 - May 1, 2005, Madison, WI, U.S.A. Presented by Fredrick Olness.}}

\begin{abstract}
Analysis of semi-inclusive DIS hadroproduction suggests broadening
of transverse momentum distributions at small $x$ below $10^{-3}\sim10^{-2}$,
which can be modeled in the Collins-Soper-Sterman formalism by a modification
of impact parameter dependent parton densities. We investigate these
consequences for the production of electroweak bosons at the Tevatron
and the LHC. If substantial small-$x$ broadening is observed in forward
$Z^{0}$ boson production in the Tevatron \hbox{Run-2}, it will strongly
affect the predicted $q_{T}$ distributions for $W^{\pm}$ and $Z^{0}$
boson production at the LHC. 
\end{abstract}
\maketitle
\title{$q_{T}$ Uncertainties for $W$ and $Z$ Production}

\textsc{Introduction:} As we move from the Tevatron collider at 1.96~TeV
to the Large Hadron Collider (LHC) at 14~TeV, we encounter an unexplored
kinematic regime. In this regime we may discover phenomena with significant
consequences for precision measurements and searches for new physics. 

In this paper,%
\footnote{The results presented here are based on Ref.~\cite{Berge:2004nt};
refer to this reference for a detailed description of the process
and more extensive references. %
} we analyze the consequences of anomalous transverse momentum ($q_{T}$)
broadening driven by possible small-$x$ effects in $W$ and $Z$
boson production at the Tevatron and LHC. At present, the form of
the $q_{T}$ distributions of $W$, $Z$, and Higgs boson production
at the LHC is largely unknown, in part because limited experimental
data on $q_{T}$ distributions is available in Drell-Yan-like processes
at small $x$. If we turn to the crossed deep inelastic scattering
(DIS) process, $q_{T}$ broadening was observed at the HERA $ep$
collider in the small $x$ region: $x=10^{-4}\sim10^{-2}$~\cite{Nadolsky:1999kb,Nadolsky:2000ky,Adloff:1999ws,Aid:1995we}.
We use these results to predict the effects in hadron-hadron processes.
The resulting modifications in the transverse momentum distributions
may affect the measurements of the $W$ boson mass and width, as well
as the $W$ and $Z$ boson background in the search for new gauge
bosons. The $q_{T}$ broadening may also affect the detection of the
Higgs boson at the LHC by altering its $q_{T}$ distribution and the
relevant QCD background. 

\begin{figure}
\includegraphics[%
  clip,
  width=0.48\textwidth,
  height=1.6in,
  keepaspectratio]{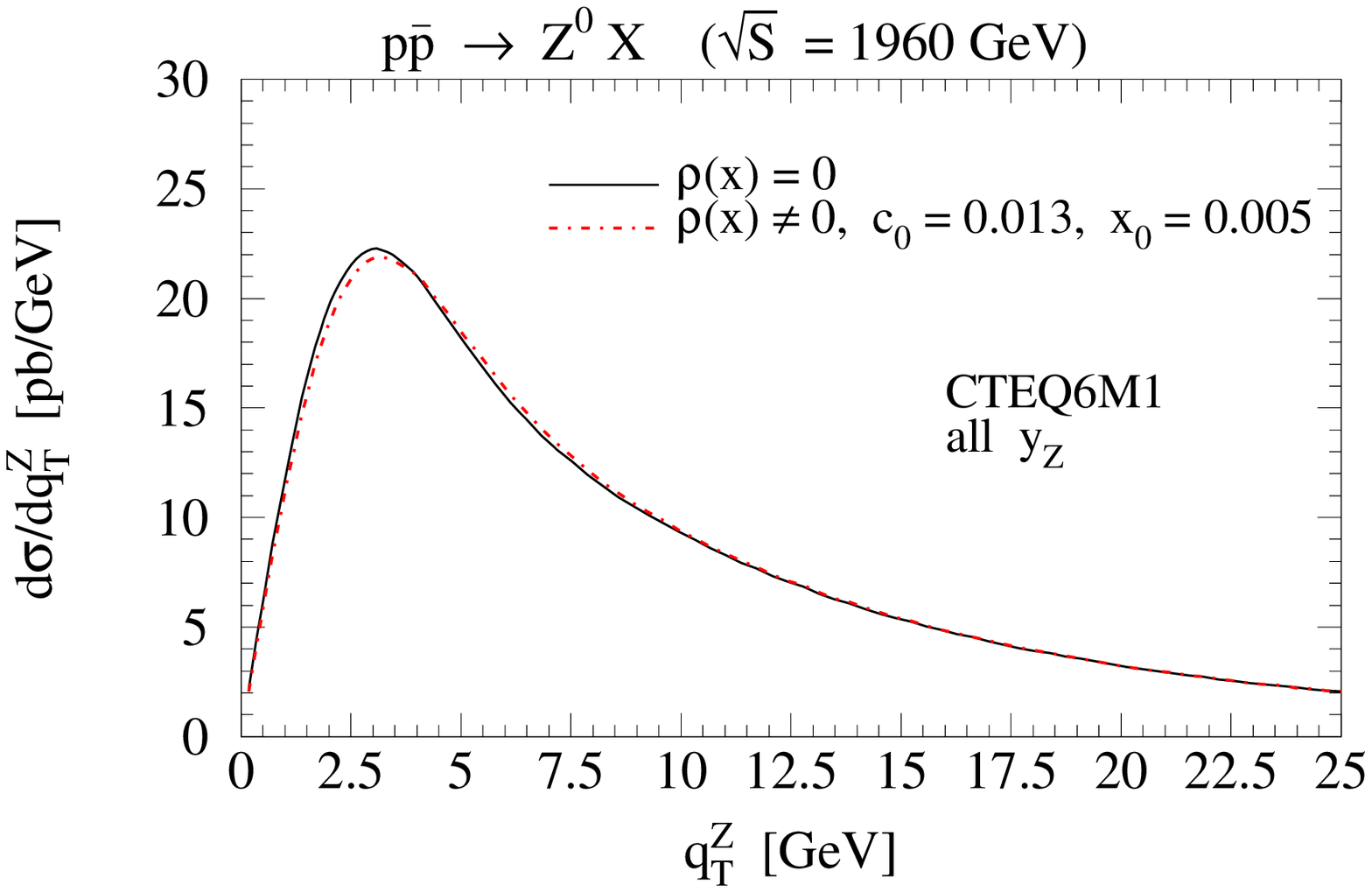}\includegraphics[%
  clip,
  width=0.48\textwidth,
  height=1.6in]{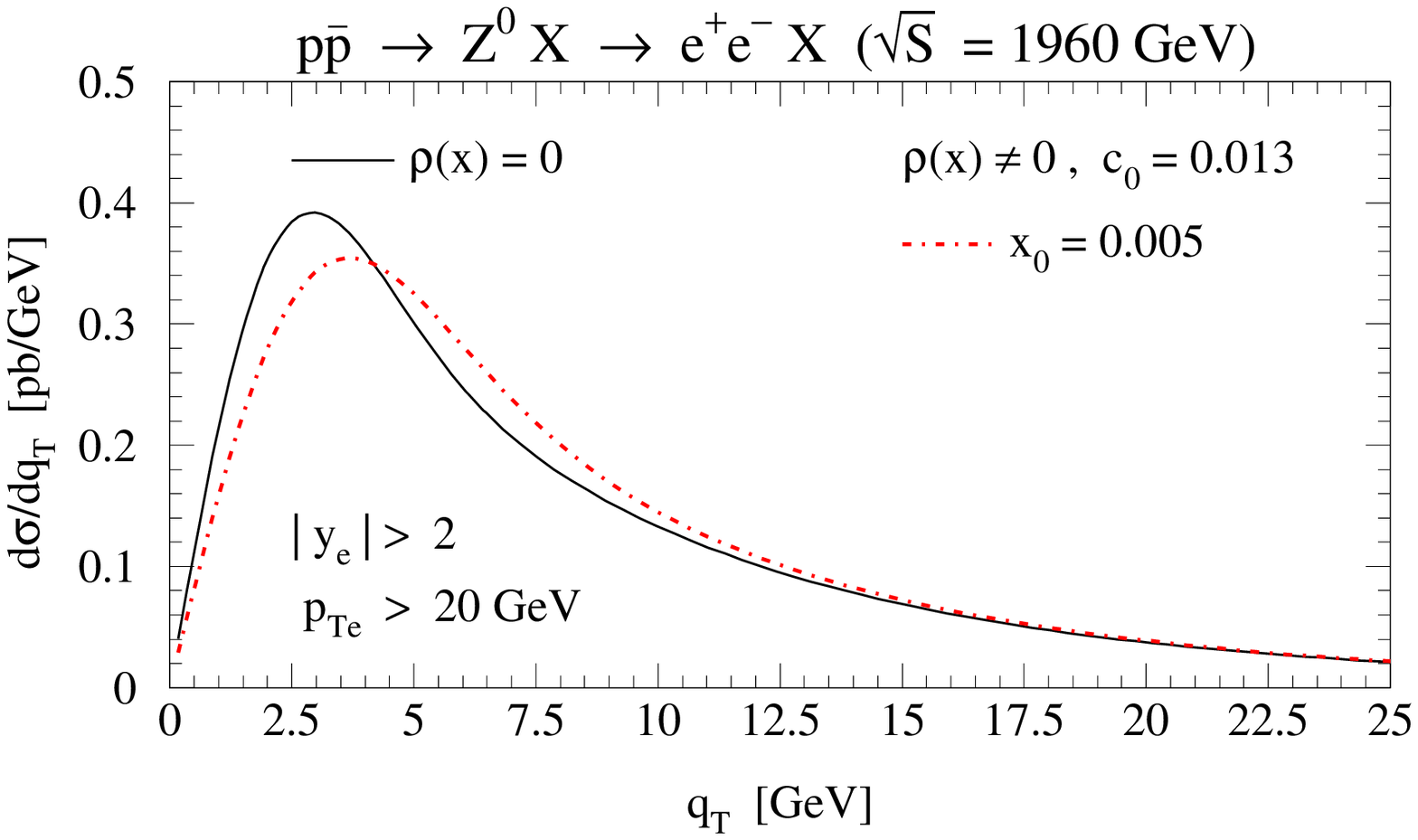}

\caption{Transverse momentum distributions of $Z$ bosons in the Tevatron
Run-2: (A)~integrated over the full range of $Z$ boson rapidity
$y$; (B)~ for events with both decay electrons registered in the
forward ($y_{e^{+}}>2,\,\, y_{e^{+}}>2$) or backward ($y_{e^{+}}<-2,\,\, y_{e^{+}}<-2$)
detector regions.  The solid curve is the standard CSS cross section,
calculated using the BLNY parametrization \cite{Landry:2002ix} of
the non-perturbative Sudakov factor. The dashed curve includes the
additional term responsible for the $q_{T}$ broadening in the small-x
region. \label{figone}}
\end{figure}

\textsc{Parameterizing the Broadening}: We now characterize the $q_{T}$
broadening which was observed in semi-inclusive DIS processes and
consider implications for the (crossed) Drell-Yan process. We examine
the resummed transverse momentum distribution for the Drell-Yan process,
following the notations of Refs.~\cite{Berge:2004nt,Berge:2004va}
\begin{equation}
\frac{d\sigma}{dydq_{T}^{2}}=\frac{\sigma_{0}}{S}\,\,\int\frac{d^{2}b}{(2\pi)^{2}}\,\, e^{-i\vec{q}_{T}\cdot\vec{b}}\,\,\widetilde{W}(b,Q,x_{A},x_{B})+Y(q_{T},Q,x_{A},x_{B}).\label{eq:w}\end{equation}
 Here $x_{A,B}\equiv Qe^{\pm y}/\sqrt{S}$, $y$ is the rapidity,
the integral is the Fourier-Bessel transform of a resummed form factor
$\widetilde{W}$ given in impact parameter ($b$) space, and $Y$
is the regular part of the fixed-order cross section ($Y$ is small
at $q_{T}\rightarrow0$). The form factor~$\widetilde{W}$ is given
by a product of a Sudakov exponent $e^{-S(b,Q)}$ and generalized
parton distributions ${\mathcal{\overline{P}}}(x,b)$:\[
\widetilde{W}(b,Q,x_{A},x_{B})\,=\frac{\pi}{S}\,\sum_{a,b}\sigma_{ab}^{(0)}\, e^{-S(b,Q)}\,{\mathcal{\overline{P}}}(x_{A},b)\,\,{\mathcal{\overline{P}}}(x_{B},b)\,.\]
 In the limit of small $b$, we can write ${\mathcal{\overline{P}}}(x,b)$
in the form: ${\mathcal{\overline{P}}}(x,b)\simeq({\mathcal{C}}\otimes f)(x,b_{0}/b)\,\, e^{-\rho(x)\, b^{2}}$,
where ${\mathcal{C}}(x,b_{0}/b)$ are coefficient functions, $f(x,\mu)$
are $k_{T}$-integrated parton distributions, and $b_{0}=2e^{-\gamma_{E}}$. 

The expressions for $\overline{\mathcal{P}}(x,b)$ differ from the
conventional form by the introduction of the term $e^{-\rho(x)\, b^{2}}$,
which will provide an additional $q_{T}$ broadening with an $x$
dependence specified by $\rho(x)$. This phenomenological characterization
of the $q_{T}$ broadening follows the corresponding analysis of the
effect observed at HERA. This $q_{T}$ broadening may approximate
$x$-dependent higher-order contributions that are not included in
the finite-order (NLO) expression for $({\mathcal{C}}\otimes f)$.
We parametrize $\rho(x)$ in the following functional form: \[
\rho(x)=c_{0}\left(\sqrt{\frac{1}{x^{2}}+\frac{1}{x_{0}^{2}}}-\frac{1}{x_{0}}\right)\]
 such that $\rho(x)\sim c_{0}/x$ for $x\ll x_{0}$, and $\rho(x)\sim0$
for $x\gg x_{0}$. This parameterization ensures that the formalism
reduces to the usual CSS form for large $x$ ($x\gg x_{0}$) and introduces
an additional source of $q_{T}$ broadening (growing as $1/x$) at
small $x$ ($x\ll x_{0}$). The parameter $c_{0}$ determines the
magnitude of the broadening for a given $x$, while $x_{0}$ specifies
the value of $x$ below which the broadening effects become important.
Based on the observed dependence $\rho(x)\sim0.013/x$ at $x\lesssim10^{-2}$
in SIDIS energy flow data, we choose $c_{0}=0.013$ and $x_{0}=0.005$
as a representative choice for our calculations.

\textsc{Kinematics:} The small-$x$ broadening occurs when one or
both longitudinal momentum fractions $x_{A,B}\approx M_{V}e^{\pm y}/\sqrt{S}$
are of order or less than $x_{0}=0.005$. Lower values of $x_{A}$
can be reached at the price of pushing $x_{B}$ closer to unity, and
vice versa; hence, we anticipate this effect will be enhanced as we
move to either large boson rapidity $|y|$, or to small $M_{V}/\sqrt{S}$.
The rate in the forward $\left|y\right|$ region is suppressed by
the decreasing parton densities at $x\rightarrow1$. 

\begin{figure}
\includegraphics[%
  clip,
  width=0.48\columnwidth,
  height=2.7in]{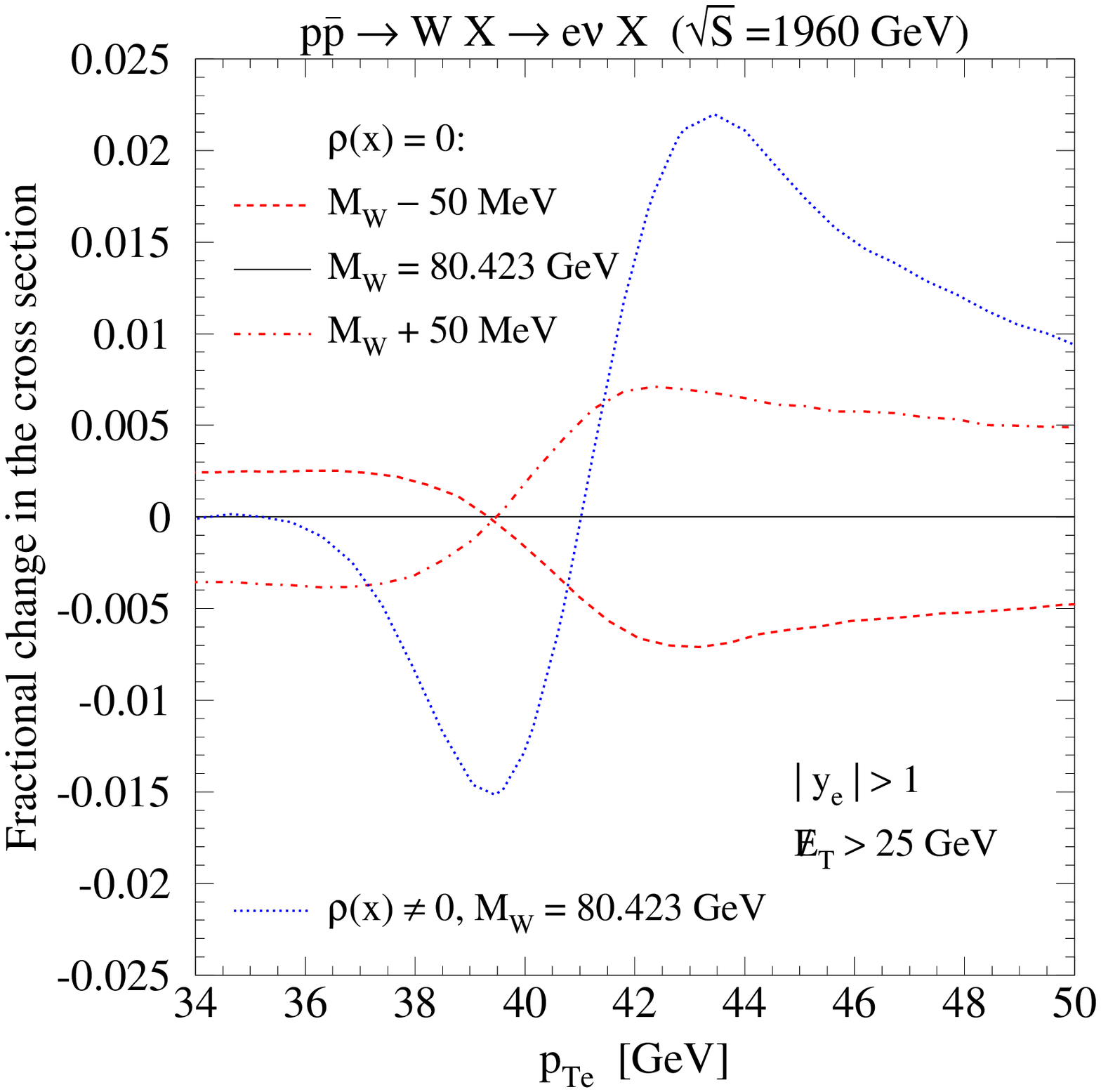}\includegraphics[%
  width=0.48\columnwidth,
  height=2.7in]{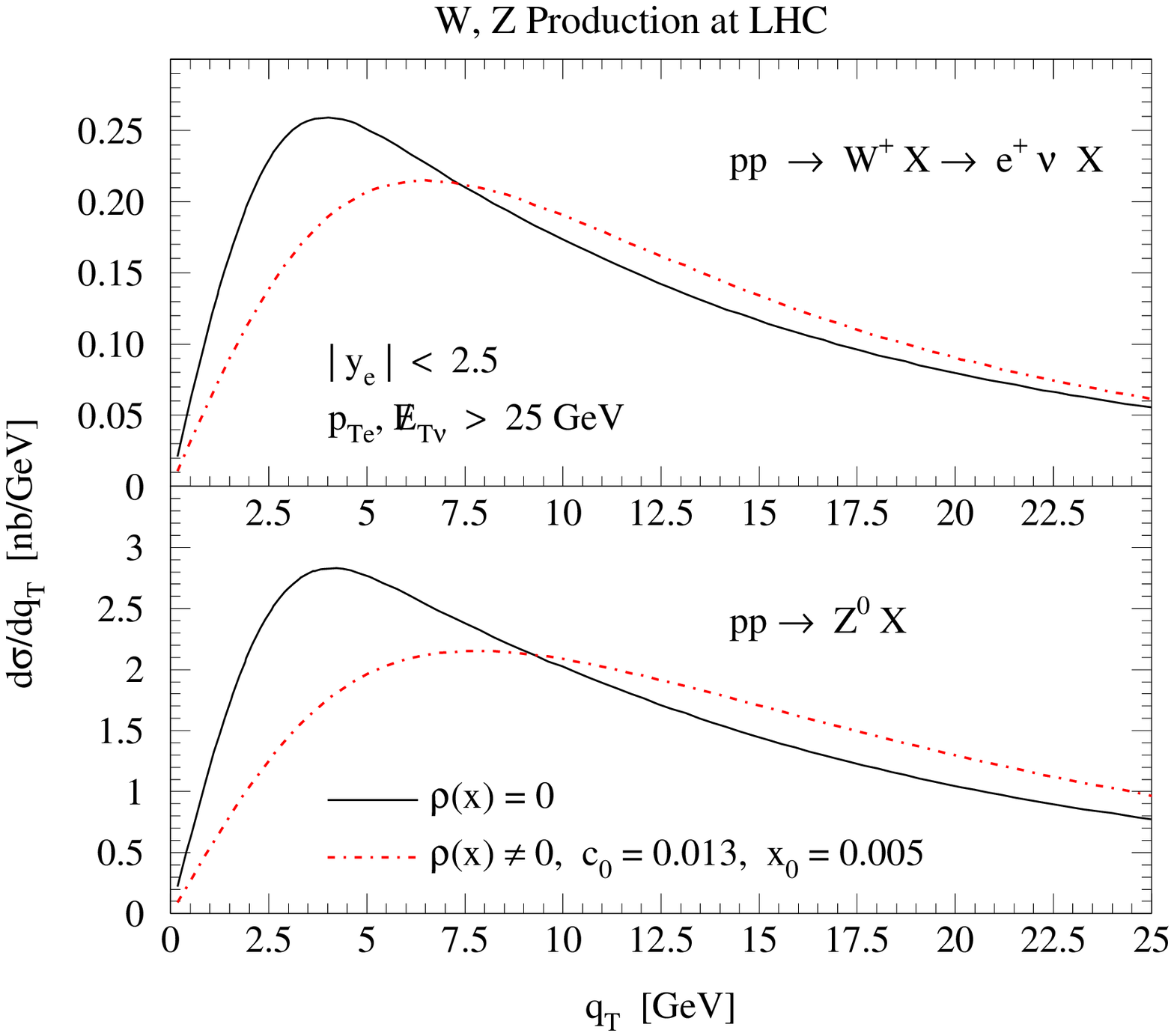}

\caption{A) The fractional difference in the distribution $d\sigma/dp_{Te}$
for the forward-rapidity sample of electrons ($|y_{e}|>1$) at the
Tevatron. B) Transverse momentum distributions of (i) $W^{+}$ bosons
and (ii) $Z^{0}$ bosons at the Large Hadron Collider. \label{figtwo}}
\end{figure}

\textsc{Z Bosons at the Tevatron:} The broadening may be observed
in the di-lepton channel in $Z$ boson production in the Tevatron
Run-2. The dominant contributions come from $x\sim M_{Z}/\sqrt{S}\sim0.046\gg x_{0}$,
where the broadening function $\rho(x)$ is negligible. Consequently,
the strategy here is to exclude contributions from the central-rapidity
$Z$ bosons, which are almost insensitive to the broadening. If no
distinction between the central and forward $Z$ bosons is made (e.g.,
as in the Run-1 analysis), the small-$x$ broadening contributes at
or below the level of the other uncertainties in the resummed form
factor. Fig.~\ref{figone}(A) shows the $Z$ boson distribution $d\sigma/dq_{T}$,
integrated over the $Z$ boson rapidity $y$ without selection cuts
on the decay leptons. The cross section with the broadening term (dashed
line) essentially coincides with the cross section without such a
term (solid line). 

In contrast, the small-$x$ broadening visibly modifies $d\sigma/dq_{T}$
at forward rapidities ($|y|<2$), where one of the initial-state partons
carries a small momentum fraction~($x\lsim0.005$). Fig.~\ref{figtwo}(B)
displays the cross sections with the acceptance cuts $y_{e^{\pm}}\,>2$
or $y_{e^{\pm}}\,<-2$ simultaneously imposed on both decay leptons.
The cuts exclude central $Z$ contributions and retain a fairly large
cross section ($\approx3.4$ pb), most of which falls within the experimental
acceptance region. Run-2 can discriminate between the curves in Fig.~\ref{figone}(B)
given the improved acceptance and increased luminosity of the upgraded
Tevatron collider; this result will have important implications for
the $W$ boson measurements, as we illustrate in the following section. 

\textsc{$M_{W}$ Measurement:} The $q_{T}$ distribution of the $W$
boson is important, as this influences the extraction of $M_{W}$
from the distribution $d\sigma/dp_{Te}$ over the transverse momentum
$p_{Te}$ of the decay charged leptons. The distribution $d\sigma/dp_{Te}$
exhibits the typical Jacobian peak located at $p_{Te}\sim\nolinebreak M_{W}/2\approx40$
GeV. To better visualize percent-level changes in $d\sigma/dp_{Te}$
associated with the broadening, we plot in Fig.~\ref{figtwo}(A)
the fractional difference $\left(d\sigma^{mod}/dp_{Te}\right)/\left(d\sigma^{std}/dp_{Te}\right)-1$
of the {}``modified'' (\emph{mod}) and {}``standard'' (\emph{std})
theory cross sections. The broadening of $d\sigma/dq_{T}$ shifts
the Jacobian peak in the positive direction. At $|y_{e}|>1$, the
small-$x$ broadening is large and exceeds the other theoretical uncertainties,
and is comparable with a variation of $M_{W}$ by more than 50~MeV.
For $|y_{e}|<1$ (not shown), the effect is comparable with a variation
of $M_{W}$ by \textasciitilde{}20~MeV. In either case, if these
effects are present, they must be taken into account for precision
measurements. 

\textsc{W \& Z Bosons at the LHC:} At the LHC, the small-$x$ broadening
can be observed in $W$ and $Z$ boson production at all rapidities
because $x<0.005$ for all $y$. Fig.~\ref{figtwo}(B-i) displays
the $q_{T}$ shift for the production of $W$ bosons with experimental
cuts for ATLAS. The shift is slightly larger in $W^{+}$ boson production
as compared to $W^{-}$ boson production (not shown) because of the
flatter $y$ distribution for $W^{+}$ bosons. The observed $q_{T}$
broadening propagates into the leptonic transverse mass and transverse
momentum distributions. The $q_{T}$ shift for the production of $Z$
bosons is comparable and displayed in Fig.~\ref{figtwo}(B-ii). A
measurement of the rapidity dependence of $q_{T}$ distributions at
the LHC will test for the presence of such effects.

\textsc{Conclusions:} For the choice of parameters extracted from
the fit to the HERA data, the $q_{T}$ broadening may be discovered
via the analysis of forward produced $Z$ bosons at the Tevatron Run-2.
Then the $q_{T}$ broadening will shift the measured $W$ boson mass
in the $p_{Te}$ method by $\sim20$ MeV in the central region ($\left|y_{e}\right|<1$),
and more than $50$ MeV in the forward region ($\left|y_{e}\right|>1$).
At the LHC, these effects produce a much harder $q_{T}$ distribution
for $W$ and $Z$ bosons and have important implications for the measurement
of the $W$ boson mass. 


\bibliographystyle{aipproc}
\bibliography{bibolness3a}

\end{document}